\documentclass[11pt]{article}

\usepackage{graphicx}
\usepackage{amsmath}
\usepackage{amssymb}
\usepackage{booktabs}
\usepackage{multirow}
\usepackage{tabularx}
\usepackage{float}
\usepackage{listings}
\usepackage{courier}
\usepackage{amsthm}
\usepackage{courier}
\usepackage{tcolorbox}
\usepackage{mdframed}
\usepackage{xcolor}
\usepackage{colortbl}
\usepackage{caption}
\tcbuselibrary{breakable}

\usepackage{acl}
\usepackage{times}
\usepackage{latexsym}
\newcounter{listingnum}
\renewcommand{\thelistingnum}{\arabic{listingnum}}

\tcbuselibrary{breakable, skins}

\newcommand{\ABListing}[3]{%
  \refstepcounter{listingnum}\label{#1}%
  \begin{tcolorbox}[
    width=0.88\textwidth,
    enhanced,
    breakable,
    colback=white,
    colframe=red!75!black,
    boxrule=0.9pt,
    arc=3pt,
    left=6pt,right=6pt,top=5pt,bottom=6pt,
    colbacktitle=red!75!black,
    coltitle=white,
    fonttitle=\bfseries,
    toptitle=3pt,
    bottomtitle=3pt,
    title={A/B Human Evaluation (#2)},
  ]
  \ttfamily\small
  #3
  \end{tcolorbox}
  
}
\usepackage[T1]{fontenc}

\usepackage[utf8]{inputenc}

\usepackage{microtype}

\usepackage{inconsolata}

\usepackage{graphicx}

\title{ES4R: Speech Encoding Based on Prepositive Affective Modeling for Empathetic Response Generation}

\author{
    Zhuoyue Gao, 
    Xiaohui Wang,
    Xiaocui Yang,
    Wen Zhang,
    \\
    \textbf{
    Daling Wang\thanks{\ \ Corresponding author.},
    Shi Feng,
    Yifei Zhang} \\
    \\ 
    Northeastern University, China\\
    \small \ttfamily{
    {gaozy5@mails.neu.edu.cn}, 
    {wangxh10@mails.neu.edu.cn},
    {yangxiaocui@cse.neu.edu.cn}, }\\
    \small \ttfamily{
    2401967@stu.neu.edu.cn, \{wangdaling, fengshi, zhangyifei1\}@cse.neu.edu.cn
     }
    }

\begin{document}
\maketitle
\begin{abstract}
Empathetic speech dialogue requires not only understanding linguistic content but also perceiving rich paralinguistic information such as prosody, tone, and emotional intensity for affective understandings. Existing speech-to-speech large language models either rely on ASR transcription or use encoders to extract latent representations, often weakening affective information and contextual coherence in multi-turn dialogues. To address this, we propose \textbf{ES4R}, a framework for speech-based empathetic response generation. Our core innovation lies in explicitly modeling structured affective context before speech encoding, rather than relying on implicit learning by the encoder or explicit emotion supervision. Specifically, we introduce a dual-level attention mechanism to capture turn-level affective states and dialogue-level affective dynamics. The resulting affective representations are then integrated with textual semantics through speech-guided cross-modal attention to generate empathetic responses. For speech output, we employ energy-based strategy selection and style fusion to achieve empathetic speech synthesis. ES4R consistently outperforms strong baselines in both automatic and human evaluations and remains robust across different LLM backbones. 
\end{abstract}
\begin{figure}[t]
    \centering
    \includegraphics[width=\linewidth]{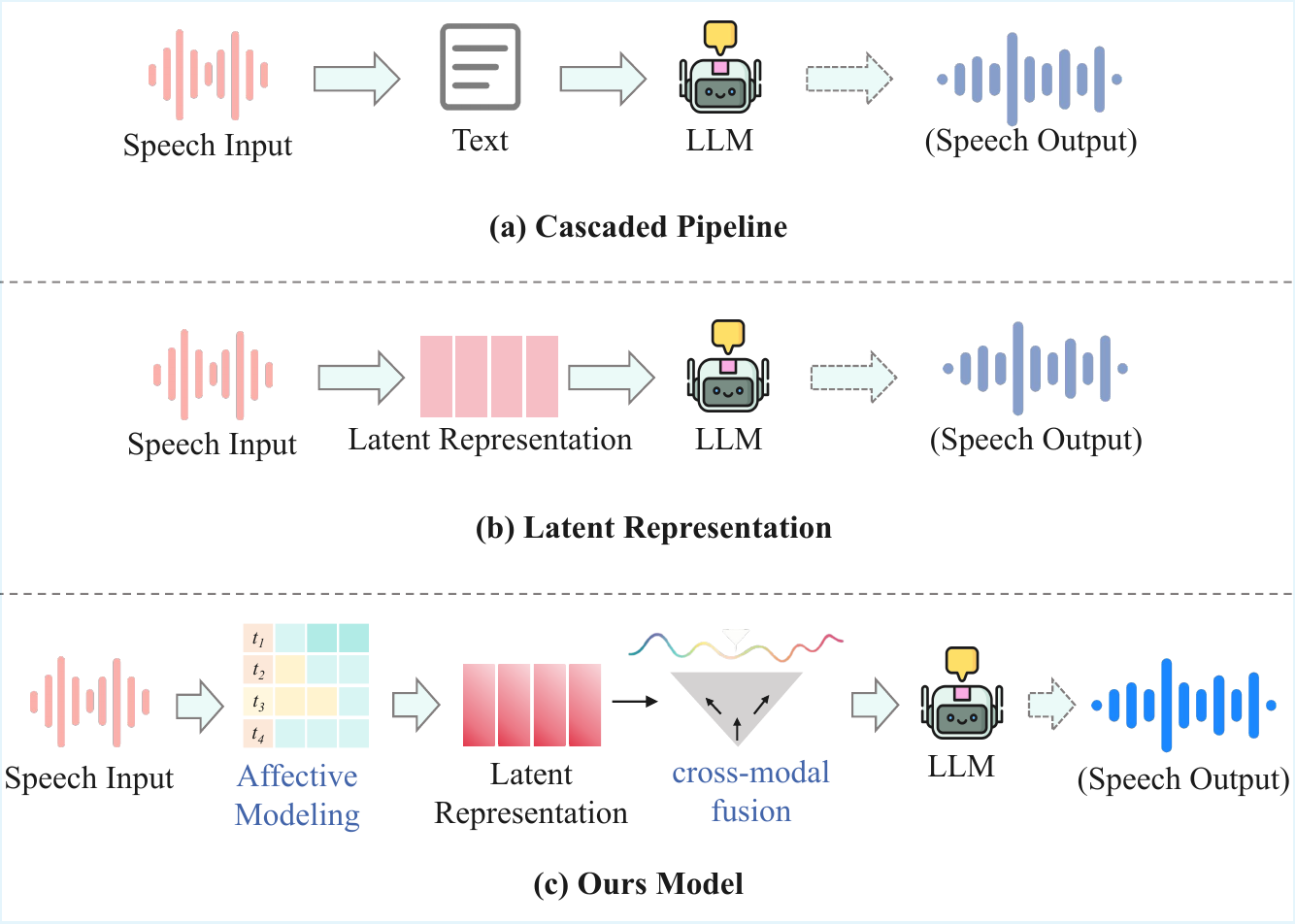}
    \caption{Comparison of speech dialogue system architectures. (a) Cascaded pipeline through ASR. (b) End-to-end latent representations. (c) Ours model.}
    \label{fig1}
\end{figure}
\section{Introduction}
Empathetic Response Generation (ERG) aims to enhance dialogue system’s understanding of users’ emotions and produce appropriate responses~\cite{keskin2014empathy,wang2023prophetic}. Early studies mainly focused on text-based empathetic generation~\cite{cao-etal-2025-tool,huangfu-etal-2025-non}, yet in real--world interactions, users often express emotions naturally through multiple modalities. Recently, multimodal empathetic dialogue systems have gained increasing attention~\cite{zhang2025towards}, integrating vision~\cite{wu-etal-2025-traits}, speech~\cite{wang-etal-2024-blsp}, and text to enhance affective understanding. 
Among these modalities, speech plays a particularly important role. Beyond semantic information, it conveys rich affective and paralinguistic information~\cite{schuller2013interspeech}, such as prosody, speaking rate, rhythm and voice quality, that largely determine not only what is communicated but also how it is expressed. Meanwhile, many empathetic speech systems rely on explicit emotion annotations or discrete emotion categories for supervision \cite{yan2024talk}, which is costly \cite{welivita2021large} and struggles to capture the continuous and subtle nature of emotions \cite{russell1980circumplex}. Therefore, effectively modeling and preserving affective information in speech is a core challenge for building high-fidelity empathetic dialogue systems.

With the development of Speech Large Language Models (SLLMs)~\cite{wang-etal-2024-blsp,wang2025opens2s,yang2025large,cui2025recent}, current speech dialogue systems typically follow two paradigms for processing speech inputs, as shown in Fig.~\ref{fig1}. Specifically, (a) \textbf{Cascaded pipeline}: speech is first transcribed into text via automatic speech recognition (ASR) and then fed into a large language model (LLM) (e.g.,~\cite{huang2024audiogpt,shen2023hugginggpt}). This paradigm struggles to preserve affective information in acoustic information during transcription, thereby limiting the model’s ability to perceive the speaker’s true emotional state. (b) \textbf{Latent representation}: This methods employ speech encoders (e.g., Whisper~\cite{radford2023robust}, HuBERT \cite{hsu2021hubert}) to convert raw speech into frame-level representations, which are subsequently aligned with LLM embeddings through modality adapters~\cite{hono-etal-2024-integrating,10389732,hu2024wavllm}. However, by relying on general-purpose encoders and performing early compression of speech information \cite{schuller2011interspeech,sainath2015learning} and only performing simple concatenation or alternating embedding of the context after encoding, they risk weakening affective and paralinguistic information critical for empathetic response generation.

To address these issues, we propose \textbf{ES4R} (Empathetic Speech for Response), a framework for speech-based empathetic response generation. Our core insight is that speech affective information should be explicitly modeled before speech encoding, rather than being implicitly learned by the encoder.  ES4R follows a three stage framework including empathetic understanding, empathetic generation, and speech synthesis. As shown in Fig.~\ref{fig1}(c), in the understanding stage, we construct structured affective context modeling on speech inputs prior to encoding them into latent representations. Specifically, intra-turn\footnote{We define each "turn" as one complete speaker-listener exchange (i.e., one question-answer pair).} attention learns affective expression state of a single turn, while inter-turn attention models contextual affective dynamics, producing an enhanced affective context representations for subsequent modules. In the generation stage, the system performs speech-guided cross-modal fusion based on these representations to activate affective-relevant semantic information and generate appropriate empathetic textual responses. In the synthesis stage, the system leverages energy trajectory across the dialogue history and performs empathetic response strategy selection selection to dynamically adjust synthesis parameters, producing resonant speech replies.
In summary, our main contributions are as follows:

\begin{itemize}
    \item We propose ES4R, a framework for speech-based empathetic response generation that integrates empathetic understanding, generation, and speech synthesis, achieving holistic performance optimization from speech input to empathetic speech responses.
\end{itemize}
\begin{itemize}
     \item We are the first to explicitly model structured affective context before speech encoding, capturing intra-turn affective states and inter-turn affective dynamics, thereby mitigating the weakening of affective information caused by early compression and post-encoding fusion.
\end{itemize}
\begin{itemize}
    \item We conduct extensive experiments on the \textbf{AvaMERG} dataset, demonstrating the effectiveness and robustness of ES4R and showing that speech encoding based on prepositive affective modeling substantially improves empathetic understanding and response quality.
\end{itemize}

\begin{figure*}[t]
    \centering
    \includegraphics[width=\linewidth]{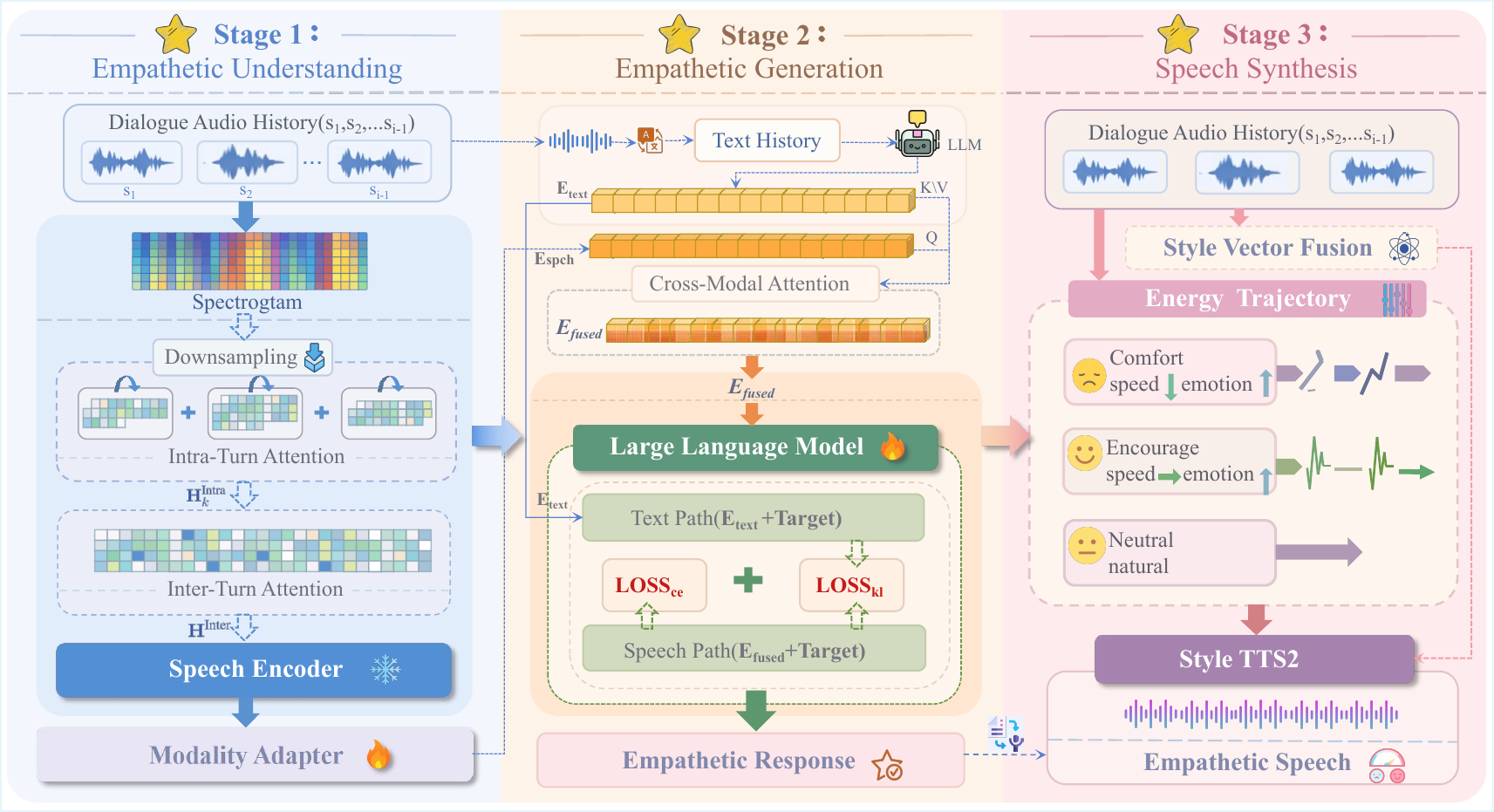}
    \caption{The overall framework of \textbf{ES4R}. \textbf{Stage 1}: Intra-Turn and Inter-Turn attention are used to prepositive affective modeling. \textbf{Stage 2}: Speech-guided cross-modal fusion is used for empathetic response generation. \textbf{Stage 3}: Energy-based strategy selection and style fusion are applied for speech synthesis.}
    \label{fig2}
\end{figure*}

\section{Related Works}
\subsection{Speech-based Dialogue Systems}
With the development of Speech Large Language Models (Speech Large Language Models, SLLMs), dialogue systems are now able to directly process audio inputs \cite{zhang-etal-2023-speechgpt,hu-etal-2024-wavllm}. Existing studies generally follow two main paradigms. Cascaded pipelines first convert speech into text via automatic speech recognition (ASR) and then feed the transcribed text into an LLM~\cite{huang2024audiogpt,shen2023hugginggpt}. This sequential processing discards paralinguistic features, such as prosody, speaking rate, and voice quality, during transcription, thereby preventing the model from perceiving emotional expression in speech. End-to-end methods~\cite{tang2023salmonn,chu2024qwen2} employ speech encoders to directly extract speech representations and align them with the LLM embedding space through modality adapters~\cite{hono-etal-2024-integrating,6638947,10389732,mitsui-etal-2024-pslm}. Such early-stage compression and abstraction may attenuate fine-grained acoustic information, which often carries emotional variations.
In multi-turn dialogue scenarios, existing SLLM architectures still face challenges in modeling temporal dependencies across turns \cite{yang-etal-2024-air,NEURIPS2024_681fe4ec}. For example, SALMONN segments dialogues into independent windows with limited interaction between windows~\cite{tang2023salmonn}, while OpenS2S incorporates dialogue history by sequentially concatenating speech and text embeddings~\cite{wang2025opens2s}. These shallow fusion strategies lack explicit mechanisms to capture evolving affective dynamics throughout the dialogue history.
To address these limitations, our framework adopts prepositive affective modeling design to enhance speech context representations before generic speech encoding and modality alignment. By modeling both intra-turn affective states and inter-turn emotional dynamics, this design effectively enhances affective information and enables structured context understanding.

\subsection{Empathetic Speech Response Generation}
Recent end-to-end empathetic spoken dialogue systems often rely on explicit emotion supervision or large-scale annotated resources \cite{yan-etal-2024-talk,ChenCKLPTSH24}. BLSP-Emo aligns emotion using SER datasets~\cite{wang-etal-2024-blsp}, and OSUM-EChat builds the EChat-200K dataset with multi-dimensional labels including emotion~\cite{geng2025osum}. Although effective, these approaches face high annotation costs, limited ability of discrete emotion categories to capture continuous emotional variations, and subjectivity in labeling \cite{chen-etal-2024-emoknob,jeon-etal-2025-prompt,hu-etal-2025-chain}. 
In contrast, we present a streamlined speech-driven approach that does not rely on explicit emotion recognition or discrete emotion supervision. We construct a structured affective context representations directly from the speech input in Empathetic Understanding. This representations is then leveraged in Empathetic Generation via speech-guided cross-modal fusion to guide semantic understanding, and further applied in Speech Synthesis, where energy-based empathetic strategy selection and style fusion are employed to generate coherent and natural empathetic spoken responses.

\section{Method}
\subsection{Task Definition}
Given dialogue audio history 
$\mathcal{S} = \{s_1, ..., s_{i-1}\}$ where each $s_k$ is a speech 
turn, our goal is to generate empathetic responses in both text and 
speech forms. The task involves three stages: empathetic 
understanding \textbf{(3.3)} from speech, empathetic response generation \textbf{(3.4)}, 
and speech synthesis \textbf{(3.5)}. The input is speech 
dialogue history, and the output are empathetic text and speech responses.
\subsection{Architecture}
ES4R comprises the following components: a speech encoder (with parameters $\psi$, augmented with a dual-level attention module $\theta_{hier}$), and a modality adapter (with parameters $\theta$) in \textbf{Stage 1}. A cross-modal fusion module (with parameters $\theta_{cross}$) and an instruction-following LLM (with parameters $\phi$) (see Appendix \ref{appendix:LLM} for details) in \textbf{Stage 2}. An empathetic speech synthesis module (with parameters $\theta_{tts}$) based on StyleTTS2~\cite{li2023styletts} in \textbf{Stage 3}. Fig. \ref{fig2} provides an overview of ES4R.

\subsection{Empathetic Understanding}
As shown in \textbf{Stage~1} of Fig.~\ref{fig2}, given the dialogue history $\mathcal{S}=\{s_1,\ldots,s_{i-1}\}$, where $s_k$ denotes the speech input of the $k$-th turn. 
Firstly, we use the Whisper~\cite{radford2023robust} feature extractor to extract the time-frequency acoustic features for each turn:
\begin{equation}
\mathbf{X}_k = \mathrm{FE}(s_k)\in \mathbb{R}^{T_k \times d_f},
\end{equation}
where $\mathrm{FE}(\cdot)$ represents the feature extractor, $\mathbf{X}_k$ is the acoustic feature matrix, $T_k$ is the number of time frames, and $d_f$ is the feature dimension (the \textit{Spectrogram} in figure corresponds to the visualization of $\mathbf{X}_k$). 
Subsequently, we introduce downsampling module on the input side to obtain $\tilde{\mathbf{X}}_k$, controlling the sequence length.

Next, we employ dual-level attention mechanism for prepositive affective modeling of multi-turn speech context . First, \textbf{Intra-Turn Attention} models dependencies within individual turns, highlighting key affective segments:
\begin{equation}
\mathbf{H}^{\text{Intra}}_k=\mathrm{MHSA}(\tilde{\mathbf{X}}_k,\tilde{\mathbf{X}}_k,\tilde{\mathbf{X}}_k),
\end{equation}
where $\mathrm{MHSA}(\cdot)$ denotes multi-head self-attention, and $\mathbf{H}^{\text{Intra}}_k$ represents the attention output for the $k$-th turn. Subsequently, we concatenate the representations of each turn in chronological order to obtain a dialogue-level sequence:
\begin{equation}
\mathbf{H}^{\text{hist}}=\mathrm{Concat}\!\left(\mathbf{H}^{\text{Intra}}_1,\ldots,\mathbf{H}^{\text{Intra}}_{k-1}\right),
\end{equation}
where $\mathbf{H}^{\text{hist}}$ is the concatenated dialogue history representations. We then apply \textbf{Inter-Turn Attention} to model contextual dependencies and the dynamic of affective as the dialogue progresses:
\begin{equation}
\mathbf{H}^{\text{Inter}}=\mathrm{MHSA}(\mathbf{H}^{\text{hist}},\mathbf{H}^{\text{hist}},\mathbf{H}^{\text{hist}}),
\end{equation}
where $\mathbf{H}^{\text{Inter}}$ is the attention output for contextual affective dynamics.

After completing prepositive affective modeling, we feed the $\mathbf{H}^{\text{Inter}}$ into a generic speech encoder (\textit{Speech Encoder} in the figure), and map it to the LLM's embedding space through a modality adapter (\textit{Modality Adapter}), providing affective representations for next stage.

\subsection{Empathetic Generation}
As shown in \textbf{Stage~2} of Fig.~\ref{fig2}, the goal of \textbf{Stage~2} is to make the speech contextual affective representations obtained from \textbf{Stage~1} play a dominant role in semantic retrieval and fusion, allowing speech to select which historical semantics are more worthy of attention, and generating responses that better align with the current dialogue state.

Let $\mathbf{E}_{\text{text}}$ denote the embedding sequence of text history (obtained from LLM word embeddings), and $\mathbf{E}_{\text{spch}}$ denote the speech-side sequence representations (obtained from \textbf{Stage~1} through modality adapter alignment to the LLM latent space). We perform \textbf{Cross-Modal Attention} with speech as the query ($\mathrm{Q}$) and text as keys and values ($\mathrm{K}/\mathrm{V}$):
\begin{equation}
\mathbf{E}_{\text{fused}} = \text{CrossAttn}(\mathbf{E}_{\text{spch}}, \mathbf{E}_{\text{text}}, \mathbf{E}_{\text{text}})
\end{equation}
Then we obtain speech representations enhanced through semantic understanding $\mathbf{E}_{\text{fused}}$.

To learn speech understanding while preserving the LLM's text capabilities, we adopt a two-path input strategy : \emph{Text Path} takes the text dialogue history $\mathbf{E}_{\text{text}}$ and target response $\mathbf{E}_{\text{tgt}}$ as input, while \emph{Speech Path} replaces the text dialogue history with the fused speech representations $\mathbf{E}_{\text{fused}}$. And we employ PLoRA~\cite{dong2024internlm} to apply LoRA only to the fused speech part, minimizing disruption to the original text instruction capabilities. 

The training objective consists of two components. The cross-entropy loss utilizes the Speech Path to generate correct responses conditioned on speech-fused representations:
\begin{equation}
\mathcal{L}_{\text{CE}} = -\frac{1}{|\mathcal{V}|}\sum_{t\in\mathcal{V}} \log p_{\theta}(y_t \mid \mathbf{E}_{\text{fused}}, y_{<t}),
\end{equation}
where $|\mathcal{V}|$ is the number of valid target positions, $y_t$ is the ground-truth token at position $t$, $y_{<t}$ denotes all preceding tokens, and $p_{\theta}(\cdot)$ is the model's predicted probability distribution with parameters $\theta$.
The KL distillation loss uses \emph{Text Path} as a teacher to guide \emph{Speech Path}, ensuring the speech-fused model learns rich semantic distributions:
\begin{align}
\mathcal{L}_{\text{KL}} &= \frac{1}{|\mathcal{V}|} \sum_{t\in\mathcal{V}} 
\text{KL}\Big(
p_{\theta}^{\text{spch}}(y_t \mid \mathbf{E}_{\text{fused}}, y_{<t}) \,\big\|\, \notag \\
&\quad p_{\theta}^{\text{text}}(y_t \mid \mathbf{E}_{\text{text}}, y_{<t}) 
\Big),
\end{align}
where $p_{\theta}^{\text{spch}}$ and $p_{\theta}^{\text{text}}$ are the output distributions from the Speech Path and Text Path, respectively. The total loss is as follows and no weight is set:
\begin{equation}
\mathcal{L}_{\text{total}} = \mathcal{L}_{\text{CE}} +  \mathcal{L}_{\text{KL}}
\end{equation}

\subsection{Speech Synthesis}
 As shown in \textbf{Stage 3} of Fig.~\ref{fig2}, after obtaining the empathetic textual response, we extract the energy trajectory from the dialogue history audio $\{s_1,\ldots,s_{i-1}\}$, and select the appropriate empathetic strategy based on its evolution trend, adjusting the prosody control parameters ($\alpha$, $\beta$) of StyleTTS2 to generate speech with different styles. We use the energy trajectory as a lightweight and interpretable proxy for affective intensity dynamics, and map its trend to supportive speaking styles (comfort/encourage/neutral) to realize empathetic prosody adaptation without explicit emotion supervision. Specifically, we use the average energy $e_k$ (calculated as the $\ell_2$ norm of the mel-spectrogram) to quantify energy level for each turn:
\begin{equation}
\Delta_e = \frac{e_{k-1} - e_1}{k-2}.
\end{equation}
When $\Delta_e < 0$ (sustained energy decrease), the \textbf{comfort} strategy is triggered; when $\Delta_e > 0$, the \textbf{encourage} strategy is activated; otherwise, the \textbf{neutral} strategy is adopted. Accordingly, we adjust the generation hyperparameters of StyleTTS2: \textbf{comfort} uses $\alpha = 0.85, \beta = 1.2$; \textbf{encourage} uses $\alpha = 1.0, \beta = 1.1$; and \textbf{neutral} uses $\alpha = 0.95, \beta = 1.0$. Here, $\alpha$ controls phoneme duration (smaller values indicate slower speech), while $\beta$ controls the emotional expressiveness in diffusion sampling.

To prevent subtle negative expressions in the dialogue history from being "averaged out", we introduce an empathy memory weighting mechanism. Specifically, we extract a style vector $\mathbf{s}_i$ from each utterance using the StyleTTS2 style encoder (which encodes both acoustic and prosodic information), and fuse them via inverse energy weighting:
\begin{equation}
w_k = \frac{1/(e_k + \epsilon)}{\sum_{j=1}^{i-1} 1/(e_j + \epsilon)}, \qquad
\mathbf{s}_{\text{fused}} = \sum_{k=1}^{i-1} w_k \cdot \mathbf{s}_k,
\end{equation}
where $e_k$ is the average energy of the $k$-th turn, $i$ denotes the current turn index, and $\epsilon = 0.001$ is a smoothing term. This weighting scheme assigns higher importance to lower-energy turns, allowing the fused style to focus more on potential negative emotional states. Finally, the fused style vector $\mathbf{s}_{\text{fused}}$, the parameters $(\alpha, \beta)$, and the generated text are jointly fed into the StyleTTS2 sampler to produce empathetic speech responses that are consistent with the dialogue context.

\begin{table*}[t]
\centering
\begin{minipage}{\textwidth}
\centering
\begin{tabular}{lcccc}
\toprule
\textbf{Model} & \textbf{BLEU-1/2/3/4} & \textbf{B-S} & \textbf{ROU-1/2/L.} & \textbf{Dist-1/2} \\
\midrule
ASR+LLM & 0.1701/0.0801/0.0498/0.0324 & 0.8681 & 0.1826/0.0813/0.1519 & \textbf{0.0320}/\textbf{0.1978} \\
Whisper+LLM & 0.1729/0.0616/0.0326/0.0193 & 0.8724 & 0.1822/0.0259/0.1379 & 0.0185/0.0735 \\
BLSP-emo & 0.2708/0.1417/0.0821/0.0514 & 0.8688 & 0.2694/0.0770/0.2153 & 0.0219/0.1145 \\
SALMONN  & 0.2042/0.1052/0.0368/0.0254 & 0.8509 & 0.2098/0.0921/0.1426 & 0.0202/0.0908 \\
OpenS2S & 0.2690/0.1403/0.0862/0.0556 & 0.8834 & 0.2676/0.0869/0.2161 & 0.0211/0.0938 \\
LLaMA-Omni 2 & 0.2202/0.0911/0.0504/0.2984 & 0.8783 & 0.2246/0.0456/0.1734 & \underline{0.0317}/\underline{0.1571} \\
Qwen2.5-Omni & 0.2737/0.1575/0.0952/0.0683 & 0.8844 & 0.3065/0.1129/0.2460 & 0.0256/0.1394 \\
\midrule
\textbf{ES4R(Qwen)} & \textbf{0.2851}/\textbf{0.1642}/\underline{0.1095}/\underline{0.0758} & \textbf{0.9058} & \textbf{0.3317}/\textbf{0.1222}/\textbf{0.2566} & 0.0225/0.1279 \\
\textbf{ES4R(llama)} & \underline{0.2773}/\underline{0.1603}/\textbf{0.1099}/\textbf{0.0773} & \underline{0.9055} & \underline{0.3259}/\underline{0.1192}/\underline{0.2511} & 0.0244/0.1490 \\
\midrule
w/o Dual-Attn & 0.2264/0.0552/0.0184/0.0078 & 0.8798 & 0.2167/0.0114/0.1600 & 0.0169/0.1007 \\
w/o Cross-Attn & 0.2659/0.1243/0.0716/0.0448 & 0.8830 & 0.2352/0.0606/0.1832 & 0.0199/0.0892 \\
\bottomrule
\end{tabular}
\end{minipage}
\captionsetup{position=bottom} 
\caption{Automatic evaluation results on AvaMERG dataset. All models are fine-tuned on AvaMERG. w/o Dual-Attn and w/o Cross-Attn denote ablations removing the dual-level attention module and speech-guided cross-modal fusion module, respectively. \textbf{Bold} indicates the best performance and \underline{underline} indicates the second-best performance.}
\label{tab:1}
\end{table*}

\section{Experiments}
\subsection{Dataset}
We use the AvaMERG~\cite{zhang2025towards} dataset, an extension of the ED dataset. This study only uses the dialogue data from the text modality and speech modality. Detailed dataset usage instructions and examples can be found in Appendix \ref{appendix:dataset}.
\subsection{Training Details}
We use the encoder of Whisper-large-v3~\cite{radford2023robust} as the speech encoder, a convolution-based subsampler as the modality adapter, and Qwen3-8B~\cite{yang2025qwen3} as the backbone of the large language model. To validate the portability of the method, we further conduct experiments on Llama-3.1-8B-Instruct~\cite{patterson2022carbon}. Please see Appendix \ref{appendix:training} for details.

\subsection{Baselines}
To validate the effectiveness of the ES4R framework, we compare it with the following baseline models. All models were fine-tuned on the AvaMERG dataset and evaluated on the test set:

\textbf{ASR+LLM}: A cascaded system that uses Whisper-large-v3 for automatic speech recognition, and then feeds the recognized transcript into an LLM to generate textual responses.
\textbf{Whisper+LLM}: A simple end-to-end system that directly uses the output of the Whisper-large-v3 encoder as the input to the LLM, bypassing the transcription stage and without an explicit emotion recognition module.
\textbf{BLSP-Emo}~\cite{wang-etal-2024-blsp}: An end-to-end SLLM model trained with a two-stage strategy, including semantic alignment and emotion alignment.
\textbf{SALMONN}~\cite{tang2023salmonn}: A multimodal LLM that achieves speech--text alignment through a dual-encoder architecture and a window-level Q-Former.
\textbf{OpenS2S}~\cite{wang2025opens2s}: An open-source speech-to-speech dialogue system that adopts streaming joint speech-text modeling.
\textbf{LLaMA-Omni 2}~\cite{fang-etal-2025-llama}: A LLaMA-based end-to-end speech interaction model that supports low-latency speech input and output.
\textbf{Qwen2.5-Omni-7B}~\cite{xu2025qwen2}: A fully multimodal model from the Qwen series that supports multimodal understanding and generation across speech and text.

In addition to the baseline models, we propose ES4R, which is built on \textbf{Qwen3-8B}, and implement a model variant on \textbf{Llama-3.1-8B-Instruct} to validate its portability. 

\subsection{Evaluation Metrics}
\textbf{Automatic Evaluation.} We utilize Distinct-n (Dist1/2 )~\cite{li2016diversity}, BERTScore (B-S)~\cite{zhang2019bertscore} , ROUGE (ROU-1/2/L.)~\cite{fang2023eva}, and BLEU-n (BLEU-1/2/3/4)~\cite{papineni2002bleu} as the primary automatic metrics for evaluating response empathetic generation performance. Distinct-1 and Distinct-2 assess response diversity at the unigram and bigram levels, respectively. B-S leverages the pre-trained embeddings of BERT and matches words in candidate sentences with those in reference sentences based on cosine similarity. ROUGE and BLEU-n measure the similarity and relevance between generated responses and reference responses. For speech generation quality assessment, since all models produce speech response that aligns with their corresponding text responses, traditional objective metrics are not applicable. Therefore, we rely on human evaluation to assess the quality of generated speech.

\textbf{Human Evaluation.}
Human evaluation remains essential for thorough and nuanced understanding of content quality. Following prior methods~\cite{sabour2022cem}, we use A/B testing to compare baseline models with ES4R. We randomly select 100 conversation samples and compare the performance of baseline model with ES4R in pairs. We recruited three researchers specializing in emotional dialogue systems as annotators. For text response evaluation, we assess from five aspects: Topic Understanding (\textbf{Topic}), Emotion Recognition (\textbf{Emotion}), Response Specificity (\textbf{Specific}), Actionable Advice (\textbf{Action}), Empathy Depth (\textbf{Empathy}) (See in Appendix \ref{appendix:Human Text-Evaluation}).
 For speech response evaluation, we assess from three aspects: 
 Dialogue-level Speech Quality Mean Opinion Score (\textbf{DMOS-Q}), Dialogue-level Emotional Consistency Mean Opinion Score (\textbf{DMOS-C}), Dialogue-level Empathy Expressiveness Mean Opinion Score (\textbf{DMOS-E})(See in Appendix \ref{appendix:Human Speech-Evaluation}).
For the same dialogue, if our model performs better, it is annotated as Win. If it performs worse, it is annotated as Lose. If there is little difference between the two, it is annotated as Tie.

\textbf{LLM-based Evaluation.}
To comprehensively evaluate the quality of generated responses, we employ GPT-5 for automatic assessment. We assess the responses from five dimensions: \textbf{Quality, Empathy, Completeness, and Fluency}. The score for each dimension ranges from 0 to 10, and detailed evaluation guidelines can be found in Appendix \ref{appendix:LLM-Evaluation}.

\begin{table}[t]
\centering
\resizebox{\columnwidth}{!}{
\begin{tabular}{lcccc}
\toprule
\textbf{Comparisons} & \textbf{Aspects} & \textbf{Win} & \textbf{Tie} & \textbf{Lose} \\
\midrule
\multirow{5}{*}{\begin{tabular}[c]{@{}l@{}}ES4R \\ vs. OpenS2S\end{tabular}} 
& Topic & 58.3\% & 32.1\% & 9.6\% \\
& Emotion & 56.2\% & 34.5\% & 9.3\% \\
& Specific & 60.8\% & 30.2\% & 9.0\% \\
& Action & 57.4\% & 33.2\% & 9.4\% \\
& Empathy & 59.1\% & 31.6\% & 9.3\% \\
\midrule
\multirow{5}{*}{\begin{tabular}[c]{@{}l@{}}ES4R \\ vs. Llama-Omni2\end{tabular}} 
& Topic & 68.5\% & 24.8\% & 6.7\% \\
& Emotion & 66.9\% & 26.1\% & 7.0\% \\
& Specific & 70.3\% & 23.2\% & 6.5\% \\
& Action & 67.7\% & 25.4\% & 6.9\% \\
& Empathy & 69.1\% & 24.3\% & 6.6\% \\
\midrule
\multirow{5}{*}{\begin{tabular}[c]{@{}l@{}}ES4R \\ vs. w/o Cross-Attn\end{tabular}} 
& Topic & 64.2\% & 28.3\% & 7.5\% \\
& Emotion & 62.5\% & 29.8\% & 7.7\% \\
& Specific & 66.9\% & 25.7\% & 7.4\% \\
& Action & 63.6\% & 28.9\% & 7.5\% \\
& Empathy & 65.1\% & 27.5\% & 7.4\% \\
\midrule
\multirow{5}{*}{\begin{tabular}[c]{@{}l@{}}ES4R \\ vs. w/o Dual-Attn\end{tabular}} 
& Topic & 72.7\% & 21.3\% & 6.0\% \\
& Emotion & 70.5\% & 23.2\% & 6.3\% \\
& Specific & 74.9\% & 19.1\% & 6.0\% \\
& Action & 71.8\% & 22.1\% & 6.1\% \\
& Empathy & 76.3\% & 18.8\% & 4.9\% \\
\bottomrule
\end{tabular}
}
\caption{Human evaluation results comparing ES4R with baselines and ablation variants across five aspects: Topic Understanding (Topic), Emotion Recognition (Emotion), Response Specificity (Specific), Actionable Advice (Action), and Empathy Depth (Empathy).}
\label{tab:2}
\end{table}

\begin{table}[t]
\centering
\resizebox{\columnwidth}{!}{
\begin{tabular}{lcccc}
\toprule
\textbf{Comparisons} & \textbf{Aspects} & \textbf{Win} & \textbf{Tie} & \textbf{Lose} \\
\midrule
\multirow{3}{*}{\begin{tabular}[c]{@{}l@{}}ES4R \\ vs. OpenS2S\end{tabular}} 
& DMOS-E & 68.4\% & 23.7\% & 7.9\% \\
& DMOS-C & 61.2\% & 29.5\% & 9.3\% \\
& DMOS-Q & 47.8\% & 35.4\% & 16.8\% \\
\midrule
\multirow{3}{*}{\begin{tabular}[c]{@{}l@{}}ES4R \\ vs. Llama-Omni2\end{tabular}} 
& DMOS-E & 75.6\% & 18.2\% & 6.2\% \\
& DMOS-C & 69.7\% & 23.1\% & 7.2\% \\
& DMOS-Q & 48.3\% & 40.1\% & 11.6\% \\
\midrule
\multirow{3}{*}{\begin{tabular}[c]{@{}l@{}}ES4R \\ vs. w/o Emp. TTS\end{tabular}} 
& DMOS-E & 76.2\% & 18.4\% & 5.4\% \\
& DMOS-C & 67.3\% & 25.8\% & 6.9\% \\
& DMOS-Q & 58.9\% & 32.4\% & 8.7\% \\
\bottomrule
\end{tabular}
}
\caption{Human evaluation results on dialogue-level speech quality using DMOS metrics: Empathy Expressiveness (DMOS-E), Emotional Consistency (DMOS-C), and Speech Quality (DMOS-Q).}
\label{tab:3}
\end{table}

\begin{table}[t]
\centering
\resizebox{\columnwidth}{!}{
\begin{tabular}{lcccc}
\toprule
\textbf{Model} & \textbf{Qua.} & \textbf{Emp.} & \textbf{Com.} & \textbf{Flu.} \\
\midrule
ASR+LLM & 5.04 & 5.52 & 4.47 & 6.81 \\
Whisper+LLM & 5.21 & 6.54 & 5.14 & 7.42  \\
BLSP-emo & 7.07 & 7.56 & 5.87 & 8.62 \\
SALMONN & 6.94 & 7.13 & 6.37 & 7.94 \\
OpenS2S & 7.22 & 7.82 & 5.92 & 8.51 \\
LlaMA-Omni2 & 7.93 & 5.99 & 5.17 & 7.84 \\
Qwen2.5-Omni & 8 & 8.22 & 6.96 & 8.52 \\
\textbf{ES4R} & \textbf{8.27} & \textbf{8.53} & \textbf{7.65} & \textbf{8.94} \\
w/o Dual-Attn & 5.31 & 7.51 & 5.03 & 7.50 \\
w/o Cross & 5.73 & 5.90 & 6.56 & 8.34 \\
\bottomrule
\end{tabular}
}
\caption{Comparison of model performance across different metrics by LLM-based evaluation}
\label{tab:4}
\end{table}

\subsection{Results and Analysis}
\subsubsection{Main Results}
\textbf{Automatic Evaluation.} Table~\ref{tab:1} reports automatic results. By modeling affective context before speech encoding and performing speech-driven cross-modal fusion, ES4R outperforms most baselines on the majority of metrics, indicating stronger contextual understanding and empathetic expression. On semantic metrics (BLEU/ROUGE/BERTScore), ES4R achieves the best or competitive performance, suggesting improved semantic consistency. Results are stable across two LLM backbones (Qwen and Llama), demonstrating good robustness.

ES4R is slightly lower on Dist-n than some baselines. We attribute this to: (i) cascaded systems may introduce more output randomness due to error accumulation in ASR and LLM, which can inflate diversity without improving correctness; and (ii) some baselines leverage large-scale external data and further fine-tune on AvaMERG, while ES4R is trained only on AvaMERG, prioritizing semantic and affective reliability over lexical diversity.

\textbf{Human Evaluation.} We compare against two mainstream SLLMs. As shown in Table~\ref{tab:2}, ES4R wins across all text dimensions. Ablations confirm key components: removing dual-level attention (w/o Dual-Attn) reduces empathy wins to 76.3\%, highlighting the importance of pre-encoding affect modeling; removing cross-modal attention (w/o Cross-Attn) reduces wins to 65.1\%, validating speech-guided semantic grounding. For speech responses (Table~\ref{tab:3}), ES4R excels in emotional expressiveness and consistency. Although streaming baselines have advantages in fluency/latency, ES4R achieves stronger expressiveness with comparable overall quality. Removing empathetic TTS (w/o Emp.~TTS) drops DMOS-E wins to 76.2\%, showing the synthesis module is critical.

\textbf{LLM-based Evaluation.} Table~\ref{tab:4} shows ES4R consistently outperforms baselines. The empathy gains mainly come from pre-encoding affect modeling, enabling more accurate affect perception. Improvements in quality and completeness are driven by cross-modal alignment between speech affect and textual semantics, reducing deviations from incorrect emotion judgments (see details in Appendix~\ref{appendix:Ablation case}). Fluency benefits from the dual-path training strategy, which preserves text generation while strengthening speech understanding.

\begin{table}[t]
\centering
\footnotesize
\setlength{\tabcolsep}{3pt}
\begin{tabular}{l|rrrrr}
\toprule
\textbf{Model} & \textbf{ROU-L} & \textbf{B-S} & \textbf{BLEU-1} & \textbf{Dist-2} & \textbf{Emp.} \\
\midrule
\textbf{ES4R}           & \textbf{0.099} & \textbf{0.841} & \textbf{0.103} & 0.636 & \textbf{7.93} \\
OpenS2S        & 0.076          & 0.839          & 0.086          & 0.676 & 6.59 \\
LLaMA-Omni2    & 0.077          & 0.468          & 0.074          & \textbf{0.713} & 5.42 \\
\bottomrule
\end{tabular}
\caption{Zero-shot cross-dataset evaluation on MELD dataset.}
\label{tab:meld_results}
\end{table}

\begin{figure}[t]
    \centering
    \includegraphics[width=\columnwidth]{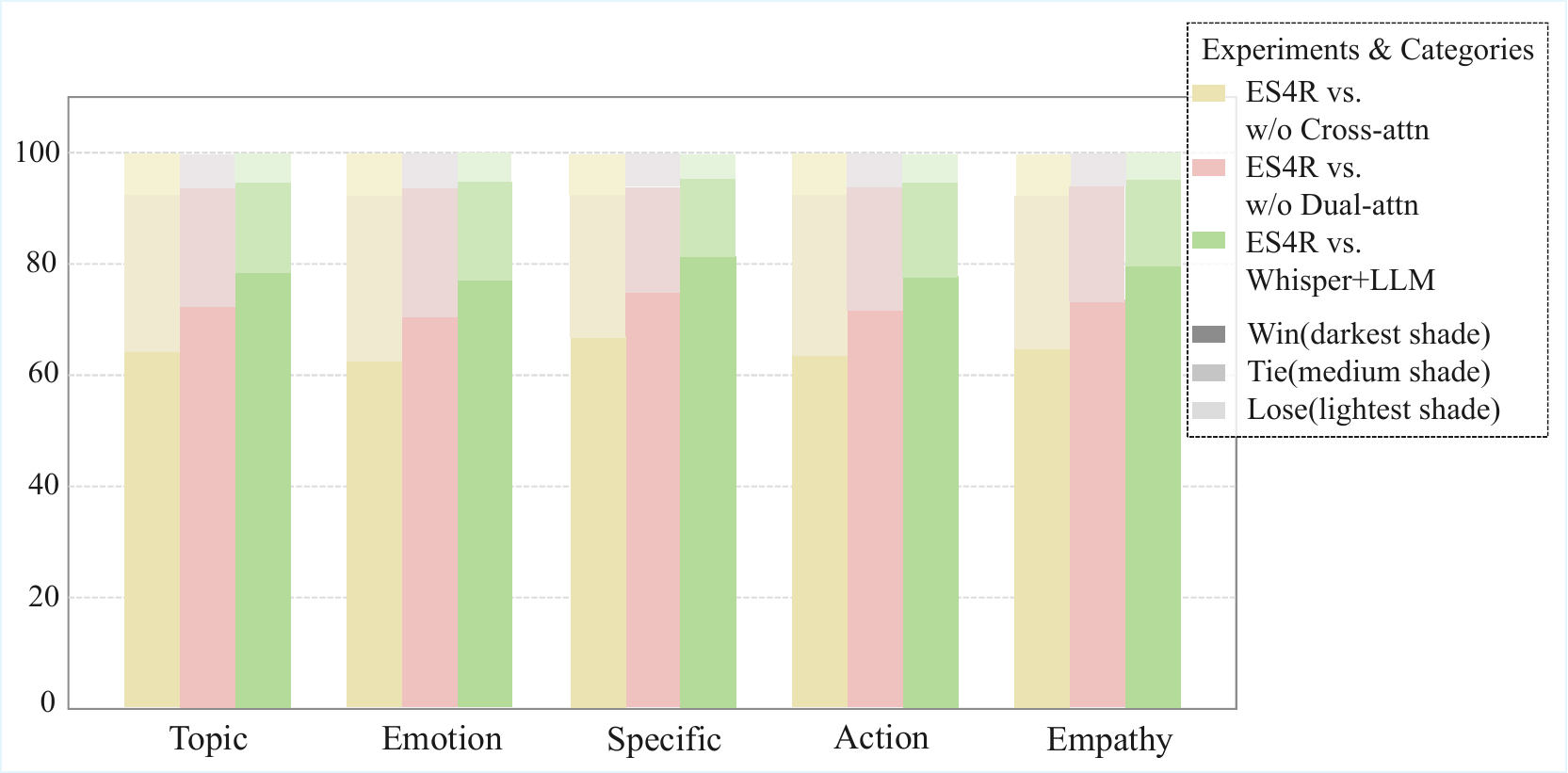}
    \caption{This figure shows multi-dimensional quality assessment across different model configurations.}
    \label{fig:3}
\end{figure}

\subsection{Cross-Dataset Generalization}

We evaluate zero-shot transfer on MELD~\cite{poria2019meld}, a multi-party dialogue dataset from TV series \textit{Friends}, without fine-tuning. As shown in Table~\ref{tab:meld_results}, ES4R outperforms baselines across most metrics (ROUGE-L: 0.099, BERTScore: 0.841, Empathy: 7.93), demonstrating strong generalization despite significant domain shift from AvaMERG's counseling dialogues. This validates that prepositive affective modeling captures transferable emotional representations.
\subsubsection{Ablation Studies}
Tables~\ref{tab:1},~\ref{tab:2},~\ref{tab:3}, and~\ref{tab:4} report the results of automatic evaluation, human evaluation (text/speech), and LLM-based assessment, together with the corresponding ablation studies. Fig.~\ref{fig:3} visualizes the performance differences across model configurations using a stacked bar chart showing win/tie/loss distributions.
In text response generation, as shown in table \ref{tab:1}, removing the dual-level attention module (w/o Dual-Attn) leads to significant performance degradation. BLEU-4 scores drop substantially. Compared to the full model, all four dimensions of LLM evaluation and five dimensions of human evaluation show significant deterioration. As shown in Fig.~\ref{fig:3}, this configuration underperforms across all evaluation aspects, demonstrating that modeling emotion context representation before encoding is a critical component of our architecture.

Removing the cross-modal attention module (w/o Cross-Attn) has a relatively smaller impact: automatic metrics show slight decreases. This variant retains reasonable capabilities in topic understanding and actionable advice, suggesting that cross-modal attention serves a complementary role when built upon high-quality emotion context representations.Compared to the Whisper+LLM baseline, ES4R demonstrates substantial advantages across all dimensions, as evidenced by the consistently higher win rates (darkest shade) in Fig.~\ref{fig:3}.


Removing the empathetic TTS module results in DMOS-E and DMOS-C win rates of 76.2\% and 67.3\%, respectively, validating its crucial role in generating expressive and emotionally appropriate speech responses. This significant performance drop demonstrates that the module effectively translates emotional understanding into perceptible prosodic features through energy trajectory modeling and strategy selection mechanisms. While we outperform the baseline in emotional expressiveness, the slightly lower DMOS-Q score reflects the inherent challenge of simultaneously maintaining both speech quality and emotional fidelity, requiring a balance between naturalness and emotional intensity.Overall, the ablation studies validate the necessity of our architectural design.

\section{Conclusion}
In this work, we propose ES4R, a framework for speech-based empathetic response generation. Unlike prior empathetic speech models that rely on explicit emotion supervision, ES4R directly constructs structured affective context representations from speech and guides the subsequent response generation process. ES4R effectively alleviates affective cue attenuation caused by early compression and improves emotional coherence in multi-turn dialogues. Experiments on the AvaMERG dataset show that ES4R significantly improves empathetic response quality in both automatic and human evaluations and remains robust across different LLM backbones, validating the effectiveness of “speech encoding based on prepositive affective modeling”. Future work will explore end-to-end streaming inference and low-latency speech generation in real interactive settings to support more natural empathetic dialogue.

\section*{Limitations}
While ES4R demonstrates strong performance, several aspects remain to be explored. First, our current empathetic speech synthesis adopts an energy-guided strategy selection with a small set of speaking styles (comfort/encourage/neutral). This design is simple and interpretable, but extending it to richer and more fine-grained controllable speaking strategies is an important direction. Second, although ES4R explicitly models affective context from speech, incorporating additional paralinguistic cues (e.g., voice quality and conversational signals) and speaker-aware personalization could further improve naturalness and adaptability. We leave these directions for future work.

\section*{Ethical Considerations}
Our research focuses on developing speech-based empathetic dialogue methods using publicly available datasets. Throughout the research process, we did not collect or process any human subject information, personal privacy data, or other sensitive information. We exclusively utilized datasets that are widely adopted in both academic and industrial communities and explicitly comply with their respective licensing agreements, with careful verification of data sources and authorization conditions. The proposed method aims to enhance empathetic speech dialogue capabilities. Beyond the general risks inherent to large language models themselves, we have not identified any additional foreseeable misuse risks or societal harms introduced by this method.

\bibliography{custom}

\appendix

\section{Instruction-following LLM}
\label{appendix:LLM}

The instruction-following large language model (LLM) serves as the core textual reasoning and response generation module in ES4R, with parameters denoted by $\phi$. 
We instantiate the LLM using \textbf{Qwen3-8B}\footnote{\url{https://huggingface.co/Qwen/Qwen3-8B}} or 
\textbf{LLaMA-3.1-8B-Instruct}\footnote{\url{https://huggingface.co/meta-llama/Llama-3.1-8B-Instruct}} 
as the base model, implemented via the HuggingFace AutoModelForCausalLM interface.

To efficiently adapt the LLM to empathetic speech interaction, we apply \textbf{Low-Rank Adaptation (LoRA)} for parameter-efficient fine-tuning. 
LoRA injects trainable low-rank matrices into the attention projection layers (i.e., $q$, $k$, $v$, and output projections), while all original pre-trained LLM parameters remain frozen during training.
In our implementation, LoRA activation is modality-aware: when the LoRA scope is set to audio, LoRA parameters are applied only to tokens corresponding to speech-derived embeddings, enabling targeted adaptation to speech-conditioned reasoning without disrupting text-only modeling.

Formally, given an instruction input sequence $x$ constructed from fused multimodal representations, the LLM generates a textual response $y$ according to:
\[
p_\phi(y \mid x) = \prod_{t=1}^{T} p_\phi(y_t \mid x, y_{<t}),
\]
where $\phi$ includes the LoRA parameters together with the frozen base LLM weights.

The LLM is trained using an autoregressive negative log-likelihood objective on response tokens. 
Additionally, a KL-divergence loss is optionally employed to align the response distributions between text-conditioned and speech-conditioned LLM outputs, encouraging modality-consistent generation.Within ES4R, the LLM performs high-level semantic and affective reasoning over fused speech--text representations and generates empathetic textual responses, which are subsequently converted into expressive speech by the empathetic speech synthesis module.

\newcommand{\speakericon}{\includegraphics[height=1.2em]{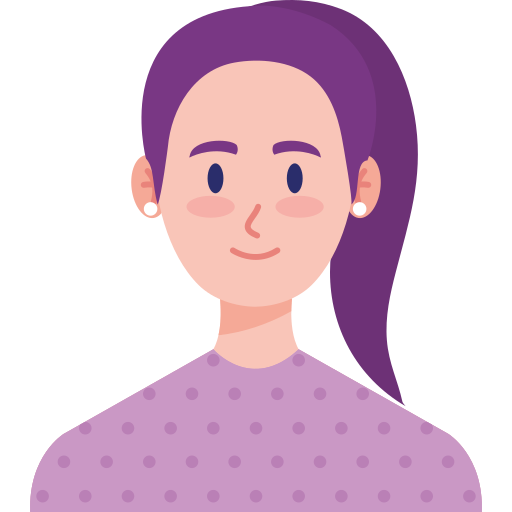}}
\newcommand{\listenericon}{\includegraphics[height=1.2em]{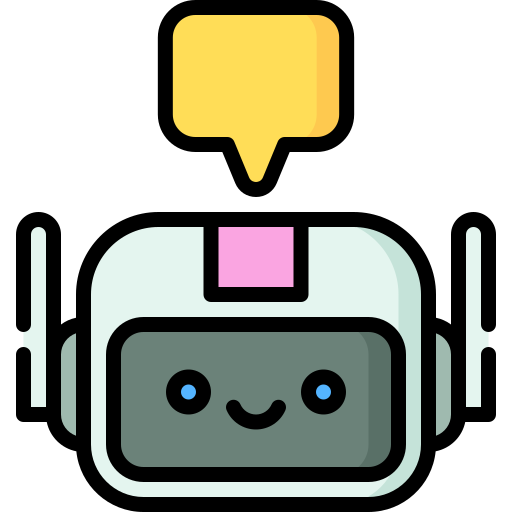}}
\begin{table*}[t]
\centering
\footnotesize
\begin{tabular}{cp{0.78\textwidth}}
\hline
\textbf{Turn} & \textbf{Content} \\
\hline
0 & \fbox{\scriptsize AUDIO} \speakericon\ \textbf{Speaker:} "I just can't shake off this feeling of being stuck in my life. It's like every day is the same and I can't see a way out." \\
  & \textcolor{black!70}{\scriptsize File: dia20002utt0\_51.wav} \\
\hline
1 & \fbox{\scriptsize AUDIO} \listenericon\ \textbf{Listener:} "It sounds really tough to feel that way. It's understandable to feel overwhelmed when life feels repetitive." \\
  & \textcolor{black!70}{\scriptsize File: dia20002utt1\_54.wav} \\
\hline
2 & \fbox{\scriptsize AUDIO} \speakericon\ \textbf{Speaker:} "Sometimes, I wonder if I'm just not trying hard enough to change things." \\
  & \textcolor{black!70}{\scriptsize File: dia20002utt2\_51.wav} \\
\hline
3 & \fbox{\scriptsize AUDIO} \listenericon\ \textbf{Listener:} "It's common to feel that way, but it doesn't mean you're not trying. Change can be slow and requires patience." \\
  & \textcolor{black!70}{\scriptsize File: dia20002utt3\_54.wav} \\
\hline
4 & \fbox{\scriptsize AUDIO} \speakericon\ \textbf{Speaker:} "I guess I just need to remind myself that small steps can lead to progress." \\
  & \textcolor{black!70}{\scriptsize File: dia20002utt4\_51.wav} \\
\hline
\rowcolor{yellow!20}
5 & \fbox{\scriptsize AUDIO} \listenericon\ \textbf{Response (Listener):} "Absolutely, recognizing that small steps matter can really shift your perspective. Every bit of progress counts." \\
  & \textcolor{black!70}{\scriptsize File: dia20002utt5\_51.wav} \\
\hline
\end{tabular}

\caption{Example multi-turn conversation from AvaMERG dataset.}
\label{tab:dataset_example}
\end{table*}

\section{Dataset Usage Instructions and Examples}
\label{appendix:dataset}
Although the AvaMERG dataset includes diverse annotations such as emotion labels, speaker characteristics, and conversation topics, our work focuses on generating empathetic responses solely from speech input. To ensure that our model learns to capture empathy directly from acoustic features and conversational context, we use only the audio modality as input, with text obtained through automatic speech transcription. This approach enables the model to process information inherent in speech signals without relying on auxiliary annotations. Table ~\ref{tab:dataset_example} illustrates a representative multi-turn empathetic conversation, where the listener provides supportive responses based on the speaker's emotional expressions. Each utterance is accompanied by its audio file, which serves as the primary input for our multimodal speech dialogue system.

We adopt a customized template for the text component because it is necessary to explicitly decompose each dialogue into independent components, including the system message, instruction, dialogue history, and response. In our architecture, audio and text are processed through parallel pathways: text is converted into embeddings via tokenization, while audio is encoded by Whisper and further processed by a dual-layer self-attention module before being fused with text through cross-attention. This design requires precise control over component boundaries, label assignment, and loss computation. Standard chat templates do not provide the level of fine-grained modular control and flexibility required for cross-modal alignment in our setting. Below, we present the dialogue templates used for Qwen3-8B(as Listing \ref{lst:qwen_format}) and Llama-3.1-8B(as Listing \ref{lst:llama_format}), respectively.

\refstepcounter{listingnum}
\label{lst:qwen_format}
\begin{tcolorbox}[
    colback=white,
    colframe=black,
    boxrule=1pt,
    arc=3pt,
    left=0pt,
    right=0pt,
    top=0pt,
    bottom=5pt,
    title={Listing \thelistingnum:Dialogue format for Qwen3-8B},
    coltitle=white,
    colbacktitle=black,
    fonttitle=\bfseries,
    toptitle=3pt,
    bottomtitle=3pt
]
\ttfamily\small
<|im\_start|>system\\
You are a helpful assistant. Your response should fulfill requests with empathy toward user's emotion tone.<|im\_end|>\\
<|im\_end|>\\
<|im\_start|>user\\
{[}speaker utterance{]}<|im\_end|>\\
<|im\_start|>assistant\\
{[}listener response{]}<|im\_end|>\\
<|im\_start|>user\\
{[}speaker utterance{]}<|im\_end|>\\
...\\
<|im\_start|>user\\
Please continue the conversation naturally as the listener<|im\_end|>\\
<|im\_start|>assistant\\
{[}target response{]}<|im\_end|>
\end{tcolorbox}

\refstepcounter{listingnum}
\label{lst:llama_format}
\begin{figure*}[t]  
\begin{tcolorbox}[
    colback=white,
    colframe=black,
    boxrule=1pt,
    arc=3pt,
    left=5pt,
    right=5pt,
    top=0pt,
    bottom=5pt,
    title={Listing \thelistingnum:Dialogue format for Llama-3.1-8B-Instruct},
    coltitle=white,
    colbacktitle=black,
    fonttitle=\bfseries,
    toptitle=3pt,
    bottomtitle=3pt
]
\ttfamily\small
\raggedright
<|begin\_of\_text|><|start\_header\_id|>system<|end\_header\_id|>\\
You are a helpful assistant. Your response should fulfill requests with empathy toward user's emotion tone.<|eot\_id|>\\
<|start\_header\_id|>user<|end\_header\_id|>\\
{[}speaker utterance{]}<|eot\_id|>\\
<|start\_header\_id|>assistant<|end\_header\_id|>\\
{[}listener response{]}<|eot\_id|>\\
<|start\_header\_id|>user<|end\_header\_id|>\\
{[}speaker utterance{]}<|eot\_id|>\\
...\\
<|start\_header\_id|>user<|end\_header\_id|>\\
Please continue the conversation naturally as the listener<|eot\_id|>\\
<|start\_header\_id|>assistant<|end\_header\_id|>\\
{[}target response{]}<|eot\_id|>
\end{tcolorbox}
\end{figure*}

\section{Training Details}
\label{appendix:training}
Table~\ref{tab:training_details} summarizes the model configuration and training hyperparameters.
We adopt a joint training strategy where the speech encoder (\textbf{Whisper-large-v3}\footnote{\url{https://huggingface.co/openai/whisper-large-v3}}) is kept frozen throughout training to preserve its robust acoustic representations.
The primary trainable components include the modality adapter and LoRA adapters applied to the LLM for parameter-efficient fine-tuning.

The modality adapter consists of three convolutional layers with kernel size 5 and stride 2, resulting in an overall downsampling factor of 8.
For LLM adaptation, we apply LoRA to the attention projection layers with rank $r=16$, scaling factor $\alpha=16$, and dropout rate 0.1.
Training is conducted for 2 epochs using the AdamW optimizer with an effective batch size of 8.
We employ BF16 mixed-precision training and DeepSpeed ZeRO-2 optimization to improve memory efficiency and training stability.

For speech synthesis, we directly utilize the pre-trained StyleTTS2 model without additional fine-tuning, leveraging its strong prosody modeling and naturalness to generate empathetic speech outputs.

\begin{table}[t]
\centering
\resizebox{\linewidth}{!}{
\begin{tabular}{l l l}
\hline
\textbf{Category} & \textbf{Parameter} & \textbf{Value} \\
\hline
\multirow{5}{*}{Audio Encoder} 
 & Backbone & Whisper-large-v3 \\
 & Sampling rate & 16 kHz \\
 & Mel bins & 128 \\
 & Trainable & No (Frozen) \\
 & Output dimension & 1280 \\
\hline
\multirow{5}{*}{Modality Adapter} 
 & Type & Convolutional subsampler \\
 & Kernel sizes & [5, 5, 5] \\
 & Stride & 2 \\
 & Downsampling factor & 8 \\
 & Hidden dimension & 512 \\
\hline
\multirow{6}{*}{LLM (Qwen / LLaMA)} 
 & Backbone & Qwen3-8B / LLaMA-3.1-8B-Instruct \\
 & Fine-tuning method & LoRA \\
 & LoRA rank $r$ & 16 \\
 & LoRA scaling $\alpha$ & 16 \\
 & LoRA dropout & 0.1 \\
 & Target modules & q\_proj, k\_proj, v\_proj, o\_proj \\
\hline
\multirow{4}{*}{Knowledge Distillation} 
 & KD loss & KL divergence \\
 & Temperature $T$ & 2 \\
 & Smoothing weight $\lambda$ & 0.5 \\
 & Applied on & Response tokens \\
\hline
\multirow{7}{*}{Training} 
 & Optimizer & AdamW \\
 & Learning rate & $5 \times 10^{-5}$ \\
 & Weight decay & 0.05 \\
 & Epochs & 2 \\
 & Effective batch size & 8 \\
 & Gradient clipping & 1.0 \\
 & Precision & BF16 \\
\hline
\multirow{4}{*}{System} 
 & Framework & PyTorch + HuggingFace \\
 & Parallelism & DeepSpeed ZeRO-2 \\
 & Checkpoint interval & 500 steps \\
 & Random seed & 42 \\
\hline
\end{tabular}
}
\caption{Model configuration and training hyperparameters.}
\label{tab:training_details}
\end{table}

\section{LLM-Evaluation on Empathetic Responses}
\label{appendix:LLM-Evaluation}
To comprehensively evaluate the quality of generated responses, we use \textbf{GPT-5} for automatic assessment. LLM-based evaluation has been shown to be highly consistent with human judgments in dialogue system evaluation. We evaluate the responses across four dimensions: Quality in Listing ~\ref{lst:quality}, Empathy in Listing ~\ref{lst:empathy}, Completeness in Listing ~\ref{lst:completeness}, and Fluency in Listing ~\ref{lst:fluency}. The score for each dimension ranges from 0 to 10, where {instruction} actually includes the complete conversation history and the last utterance of the speaker.

\refstepcounter{listingnum}
\noindent\textbf{Listing \thelistingnum:Prompt for response quality evaluation}
\label{lst:quality}

\medskip
\noindent
\ttfamily\small
Given the conversation history and the model's response. You are a helpful and precise assistant for checking the quality of the response.\\[0.5em]
<instruction>\\
\{instruction\}\\
</instruction>\\
<response>\\
\{response\}\\
</response>\\[0.5em]
Please evaluate the response with your justification having less than three sentences, and provide a score ranging from 0 to 10 after your justification. When evaluating the response, you should consider the helpfulness, harmlessness, and honesty of the response. The score should be wrapped by <score> and </score>.
\medskip
\normalfont\normalsize

\refstepcounter{listingnum}
\noindent\textbf{Listing \thelistingnum:Prompt for response empathy evaluation}
\label{lst:empathy}

\medskip
\noindent
\ttfamily\small
Given the conversation history and the model's response. You are a helpful and precise assistant for checking the empathy of the response.\\[0.5em]
<instruction>\\
\{instruction\}\\
</instruction>\\
<response>\\
\{response\}\\
</response>\\[0.5em]
Please evaluate the response with your justification having less than three sentences, and provide a score ranging from 0 to 10 after your justification. When evaluating the response, you should consider whether it shows empathy and appropriate emotional understanding. The score should be wrapped by <score> and </score>.
\medskip
\normalfont\normalsize

\refstepcounter{listingnum}
\noindent\textbf{Listing \thelistingnum:Prompt for response completeness evaluation}
\label{lst:completeness}

\medskip
\noindent
\ttfamily\small
Given the conversation history and the model's response. You are a helpful and precise assistant for checking the completeness of the response.\\[0.5em]
<instruction>\\
\{instruction\}\\
</instruction>\\
<response>\\
\{response\}\\
</response>\\[0.5em]
Please evaluate the response with your justification having less than three sentences, and provide a score ranging from 0 to 10 after your justification. When evaluating the response, you should consider whether it adequately addresses all aspects of the user's needs without omitting important information. The score should be wrapped by <score> and </score>.
\medskip
\normalfont\normalsize

\refstepcounter{listingnum}
\noindent\textbf{Listing \thelistingnum:Prompt for response fluency evaluation}
\label{lst:fluency}

\medskip
\noindent
\ttfamily\small
Given the conversation history and the model's response. You are a helpful and precise assistant for checking the fluency of the response.\\[0.5em]
<instruction>\\
\{instruction\}\\
</instruction>\\
<response>\\
\{response\}\\
</response>\\[0.5em]
Please evaluate the response with your justification having less than three sentences, and provide a score ranging from 0 to 10 after your justification. When evaluating the response, you should consider the naturalness, coherence, and linguistic quality of the generated response. The score should be wrapped by <score> and </score>.
\medskip
\normalfont\normalsize

\section{Qualitative examples}
\label{appendix:Ablation case}

\definecolor{morandi1}{RGB}{244,241,208}
\definecolor{morandi2}{RGB}{221,232,236}
\definecolor{morandi3}{RGB}{233,229,230}
\definecolor{morandi4}{RGB}{228,233,227}

\begin{table*}[t]
\centering
\setlength{\tabcolsep}{8pt}
\renewcommand{\arraystretch}{1.5}
\begin{tabular}{>{\centering\arraybackslash}p{3cm}|>{\raggedright\arraybackslash}p{13cm}}
\hline
\textbf{Model} & \textbf{Response} \\
\hline
\rowcolor{morandi2}
BLSP-emo & That's a good way to think about it. It's important to remember that progress takes time and effort, but it is possible. \\
\rowcolor{morandi3}
OpenS2S & Yes, that's right. It's easy to feel stuck, but even small steps can lead to progress. Keep going. \\
\rowcolor{morandi4}
LLaMA-Omni 2 & Exactly! And it's important to celebrate those small steps, even if they don't seem like much. \\
\rowcolor{morandi2}
Qwen2.5-Omni & That's a great mindset to have. Every small step counts and can lead to significant progress over time. \\
\hline
\rowcolor{morandi1}
\textbf{Our Model} & That's a great reminder. Sometimes it can be hard to see progress when things feel slow, but every small step counts towards making changes. \\
\hline
\rowcolor{morandi3}
w/o Dual-Attn & I'm sure you'll find something that works for you. \\
\rowcolor{morandi4}
w/o Cross-Attn & I understand. It's hard to see the light at the end of the tunnel when you're in a dark place. \\
\hline
\end{tabular}
\caption{Model Response Examples}
\label{tab:model_responses}
\end{table*}

Table~\ref{tab:model_responses} presents example model responses to the dialogue history in table \ref{tab:dataset_example} where the speaker expresses feeling lost in life and questions whether their efforts to change have been meaningful. The listener offers empathetic support, which ultimately prompts the speaker to reflect: “I think I just need to remind myself that even small steps can lead to progress.”

The baseline models exhibit varying levels of empathy. BLSP-Emo, OpenS2S, and multimodal baselines (LLaMA-Omni 2 and Qwen2.5-Omni) better understand the dialogue content and demonstrate empathy by affirming progress and celebrating small achievements. Our model generates a context-appropriate response: “That’s a great reminder. Sometimes when progress feels slow, it’s hard to see the results, but every small step is still moving you forward.” This response directly validates the speaker’s realization while providing concrete encouragement.

The ablation results show that the w/o Dual-Attn variant produces a more generic reply, while the w/o Cross-Attn variant generates: “I understand. When you are in the dark, it’s hard to see the light at the end of the tunnel.” Such a response is overly generalized and emotionally vague. These failures indicate that without dual-level attention over speech cues and cross-modal fusion, the model cannot track affective trajectories or maintain contextual coherence, leading to empathetic responses that are misaligned with the dialogue state.

\section{Human Text Evaluation on Empathetic Responses}
\label{appendix:Human Text-Evaluation}
We follow prior work~\cite{sabour2022cem} and conduct a blinded A/B preference study to compare ES4R with baseline/ablation models. We randomly sample 100 multi-turn conversations. For each sample, one response is independently generated by each model. Three researchers specializing in emotional dialogue systems independently annotate each pair on five aspects—Topic, Emotion, Specificity, Actionability, and Empathy—selecting \textit{Win/Tie/Lose} for Empath-Speech relative to the comparator (see Fig~\ref{fig:ab_criteria_listings} for the detailed criteria). Table \ref{tab:2} reports the aggregated percentages computed by counting all individual judgments across annotators and samples for each aspect, i.e., \(\text{Win/Tie/Lose} = N_{\text{win/tie/lose}}/N_{\text{valid}}\times 100\), rounded to one decimal place.

\begin{figure*}[t]
\centering

\ABListing{lst:ab_topic}{Topic Understanding / Topic}{
\textbf{Definition:} Whether the response correctly identifies and addresses the main topic and key user intent.\\
\textbf{Better response should:} A is more aligned with the user’s question/context, more relevant, and has fewer off-topic/misunderstanding issues.\\
\textbf{Tie:} Both are equally on-topic (or equally off-topic), with no clear difference.
}

\ABListing{lst:ab_emotion}{Emotion Recognition / Emotion}{
\textbf{Definition:} Whether the response accurately recognizes and responds to the user’s emotional state (e.g., anxiety, anger, sadness, happiness).\\
\textbf{Better response should:} More accurately capture the emotion and use a more appropriate tone and reaction.\\
\textbf{Tie:} Both are similarly accurate (and comparably appropriate), or both clearly misread the emotion.
}

\ABListing{lst:ab_specific}{Response Specificity / Specific}{
\textbf{Definition:} Whether the response provides concrete, situation-specific details rather than generic or templated statements.\\
\textbf{Better response should:} Be more specific, less boilerplate, and better grounded in the user’s scenario.\\
\textbf{Tie:} Both are similarly specific (or similarly generic), with no clear difference.
}

\ABListing{lst:ab_action}{Actionable Advice / Action}{
\textbf{Definition:} Whether the response provides practical, feasible, and helpful suggestions/steps.\\
\textbf{Better response should:} Offer clearer and more executable guidance that is more useful for next actions.\\
\textbf{Tie:} Both offer similar advice, or both lack actionable guidance.
}

\ABListing{lst:ab_empathy}{Empathy Depth / Empathy}{
\textbf{Definition:} The degree of emotional understanding and compassionate engagement expressed in the response.\\
\textbf{Better response should:} Better acknowledge the user’s feelings, sound more genuine/supportive, and avoid being cold, preachy, or blaming.\\
\textbf{Tie:} Both show similar empathy depth, or both are insufficient.
}

\caption{A/B human text evaluation criteria used for the five aspects reported in Table \ref{tab:2}}
\label{fig:ab_criteria_listings}
\end{figure*}

\section{Human Speech Evaluation on Empathetic Responses}
\label{appendix:Human Speech-Evaluation}
Following prior work~\cite{sabour2022cem}, we conduct a blinded A/B preference study for speech responses. We randomly sample 100 dialogues; for each dialogue, two speech outputs (Empath-Speech vs.\ a baseline/ablation) are presented in randomized order with system identities hidden. Three trained evaluators independently compare each pair on three dialogue-level aspects: speech quality (DMOS-Q), emotional consistency (DMOS-C), and empathy expressiveness (DMOS-E), using the criteria summarized in Fig.~\ref{fig:ab_speech_criteria_listings} (see Appendix~\ref{appendix:Human Speech-Evaluation} for details). For each aspect, Empath-Speech is labeled as \textit{Win} if it is preferred over the comparator, \textit{Lose} if it is worse, and \textit{Tie} if the difference is negligible. We report aggregated Win/Tie/Lose percentages by counting all individual judgments across annotators and dialogues for each aspect and rounding to one decimal place.

\begin{figure*}[t]
\centering

\ABListing{lst:ab_dmos_q}{Dialogue-level Speech Quality Mean Opinion Score / DMOS-Q}{
\textbf{Definition:} A holistic assessment of the overall quality of the generated speech.\\
\textbf{Better speech should:} Sound clearer and more intelligible, more natural (less robotic/artifact-prone), and closer to the reference speaker characteristics when applicable (e.g., timbre/style consistency).\\
\textbf{Tie:} Both are similarly good (or similarly poor) in overall speech quality, with no clear difference.
}

\ABListing{lst:ab_dmos_c}{Dialogue-level Emotional Consistency Mean Opinion Score / DMOS-C}{
\textbf{Definition:} Whether the emotional expression of the generated speech is consistent with the reference speech and the dialogue context.\\
\textbf{Better speech should:} Convey an emotion that matches the context and aligns better with the reference emotion (e.g., valence/arousal, intensity, appropriateness), without emotional mismatch.\\
\textbf{Tie:} Both are similarly consistent (or similarly inconsistent) with the reference/context.
}

\ABListing{lst:ab_dmos_e}{Dialogue-level Empathy Expressiveness Mean Opinion Score / DMOS-E}{
\textbf{Definition:} The expressiveness of empathy conveyed in the generated speech.\\
\textbf{Better speech should:} Sound more supportive and emotionally engaged (e.g., warmer prosody, appropriate emphasis/pauses), strengthening empathic intent without sounding exaggerated or insincere.\\
\textbf{Tie:} Both show similar empathy expressiveness (or both are insufficient), with no clear difference.
}

\caption{A/B human speech evaluation criteria for speech responses (DMOS-Q/C/E).}
\label{fig:ab_speech_criteria_listings}
\end{figure*}

\end{document}